\documentclass[prl,aps,twocolumn,showkeys,showpacs,amsmath,amssymb,superscriptaddress]{revtex4-1}
\usepackage{graphicx}
\usepackage{dcolumn}
\usepackage{bm}
\usepackage{epsfig}
\usepackage{subfigure}
\usepackage{color}

\begin{document} 
\title{Single to Double Hump Transition in the Equilibrium
  Distribution Function of Relativistic Particles}

\author{M. Mendoza} \email{mmendoza@ethz.ch} \affiliation{ ETH
  Z\"urich, Computational Physics for Engineering Materials, Institute
  for Building Materials, Schafmattstrasse 6, HIF, CH-8093 Z\"urich
  (Switzerland)}

\author{N. A. M. Ara\'ujo} \email{nuno@ethz.ch} \affiliation{ ETH
  Z\"urich, Computational Physics for Engineering Materials, Institute
  for Building Materials, Schafmattstrasse 6, HIF, CH-8093 Z\"urich
  (Switzerland)}

\author{S. Succi} \email{succi@iac.cnr.it} \affiliation{Istituto per
  le Applicazioni del Calcolo C.N.R., Via dei Taurini, 19 00185, Rome
  (Italy),\\and Freiburg Institute for Advanced Studies,
  Albertstrasse, 19, D-79104, Freiburg, (Germany)}

\author{H. J. Herrmann}\email{hjherrmann@ethz.ch} \affiliation{ ETH
  Z\"urich, Computational Physics for Engineering Materials, Institute
  for Building Materials, Schafmattstrasse 6, HIF, CH-8093 Z\"urich
  (Switzerland)} \affiliation{Departamento de F\'isica, Universidade
  Federal do Cear\'a, Campus do Pici, 60451-970 Fortaleza, Cear\'a,
  (Brazil)}

\date{\today}
\begin{abstract}
  We unveil a transition from single peaked to bimodal velocity
  distribution in a relativistic fluid under increasing temperature,
  in contrast with a non-relativistic gas, where only a monotonic
  broadening of the bell-shaped distribution is observed. Such
  transition results from the interplay between the raise in thermal
  energy and the constraint of maximum velocity imposed by the speed
  of light. We study the Bose-Einstein, the Fermi-Dirac, and the
  Maxwell-J\"uttner distributions, all exhibiting the same qualitative
  behavior. We characterize the nature of the transition in the
  framework of critical phenomena and show that it is either
  continuous or discontinuous, depending on the group velocity. We
  analyze the transition in one, two, and three dimensions, with
  special emphasis on two-dimensions, for which a possible experiment
  in graphene, based on the measurement of the Johnson-Nyquist noise,
  is proposed.
\end{abstract}


\maketitle

Back in 1911, F. J\"uttner derived a relativistic analogue of the
Maxwell-Boltzmann equilibrium distribution for classical
(non-relativistic) gases. To this purpose, he resorted to an entropy
minimization procedure, subject to the relativistic energy-momentum
constraints \cite{juttner}. This top-down (macro-to-micro) derivation,
left room for some debate on whether the exact form of the equilibrium
distribution for relativistic particles was the one proposed by
J\"uttner, or rather some variant thereof \cite{modjuttner1,
  modjuttner2}. Very recently, conclusive evidence for the original
form proposed by J\"uttner has been brought by numerical simulations
of fully relativistic molecular dynamics in one and two dimensions
\cite{numtest1, numtest2}. This is all but an academic exercise, since
the J\"uttner distribution is known to play a major role in the
interpretation of current and future experiments in many sectors of
modern physics, such as quark-gluon plasmas produced in heavy-ion
collisions \cite{QGP-1}, relativistic astrophysics \cite{supernova2},
distortions of the cosmic microwave background \cite{cosmic}, and
lately, possibly also in the study of electron flows in graphene
\cite{grapPRB, turbPRL}.
\begin{figure}
  \centering   
  \includegraphics[scale=0.2]{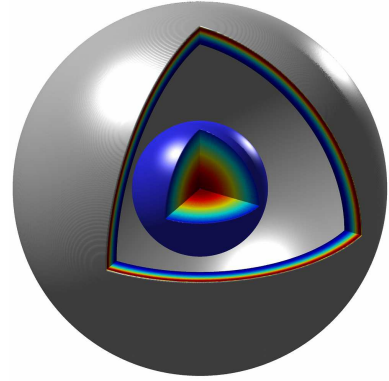}
  \caption{Three-dimensional Maxwell-J\"uttner velocity distributions,
    according to Eq.~\eqref{distribution3d}, for two different
    temperatures. The blue ($\xi=20$) and gray ($\xi=1$) isosurfaces
    stand at $1/3$ and $1/4$ of their respective maxima. Shown in the
    inner region of each isosurface, is the color gradient of the
    distribution, with red and blue colors denoting high and low
    concentration of particles, respectively.  Note that the spheres
    are plotted in velocity space, so that the maximum radius for the
    external isosurface is close to
    $|\vec{v}|=1$.\label{distribution3d}}
\end{figure}

In this Letter, we wish to call the attention on an apparently
hitherto unexplored property of the J\"uttner distribution, namely the
fact that, under the constraint of an increasing temperature
(ultra-relativistic limit $mc^2 < k_B T$), the one-dimensional
distribution develops a transition in velocity space, with the
emergence of two separate peaks, moving with opposite, non
zero-speeds.  For two and three dimensions, it generates, in velocity
space, a ring and a hollow sphere (see Fig.~\ref{distribution3d}),
respectively. This stands in sharp contrast with the way a
non-relativistic gas at rest responds to the constraint of an
increasing temperature, namely through a progressive broadening of the
Gaussian shape, which enhances the high-speed population, leaving
nonetheless the least-energetic, zero-speed, particles as the most
probable population, since this is the one best conforming to the zero
net-motion constraint. On the contrary, the transition exhibited by
the J\"uttner distribution, signals that, even in a gas at rest, the
temperature constraint can only be met by clustering most of the
particles around two oppositely moving beams, thereby {\it depleting}
the zero-speed particles in the process. The above considerations
readily generalize to the case of a moving gas, the main change being
that the two oppositely moving beams get differently populated, the
co-moving one being enhanced and the counter-moving being
correspondingly depopulated. This phenomenon is quite general, and it
might apply to a whole class of systems where physical signals are
forced to move close to the their ultimate limiting speed. In the
following, we provide mathematical details of this transition and also
discuss conditions under which it could be experimentally probed in
graphene experiments.

The probability distribution function of particle velocities in
$d$-dimensional relativistic gas is described by the following
single-particle distribution function (in natural units
($c=\hbar=e=k_B=m=1$) \cite{RelaBoltEqua,MJ1}:
\begin{equation}\label{distribution}
  f_\lambda (\vec{x},\vec{v},t) = \frac{A \gamma^{d+2}(v)}{\exp\left[\frac{1 - \vec{v}
        \cdot \vec{U}}{T}\gamma(v)\gamma(\vec{U}) - \frac{\mu}{T} \right]
    + \lambda} \quad ,
\end{equation}
where $\gamma(v) = 1/\sqrt{1-v^2}$ is the Lorentz factor, $T$ the
temperature, $\vec{U}$ the group velocity, $\vec{v}$ the velocity of
the particles, $A$ a normalization constant, $v = |\vec{v}|$, and $U =
|\vec{U}|$. The subscript $\lambda$ denotes the Fermi-Dirac ($\lambda
= 1$), the Bose-Einstein ($\lambda = -1$), and the Maxwell-J\"uttner
($\lambda = 0$) distributions, respectively.  At the moment, we will
neglect the chemical potential ($\mu = 0$).  The thermal behavior of a
relativistic gas at equilibrium is best characterized by the parameter
$\xi \equiv mc^2/k_BT = 1/T$, that is commonly used to differentiate
between the ultra-relativistic ($\xi \gg 1$) and the relativistic
($\xi \ll 1$) regimes.

\begin{figure}
  \centering   
  \includegraphics[width=\columnwidth]{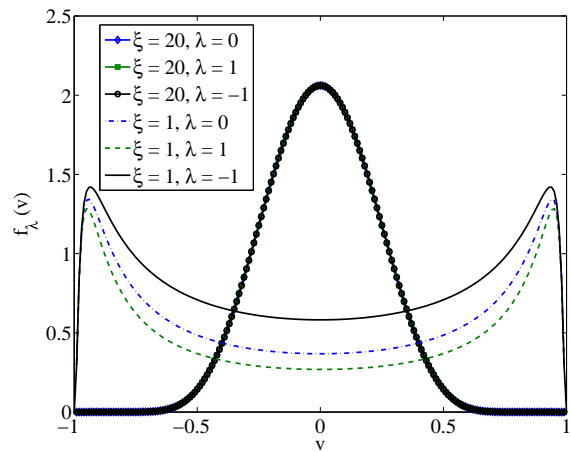}
  \caption{Bose-Einstein ($\lambda=-1$), Maxwell-J\"uttner
    ($\lambda=0$), and Fermi-Dirac ($\lambda=1$) velocity
    distributions, for $d=1$, according to Eq.~(\ref{distribution}).
    The distributions correspond to a relativistic fluid, with group
    velocity $U=0$ and two different temperatures ($T=1/\xi$), namely,
    $\xi=1$ and $\xi=20$.\label{distribution1d}}
\end{figure}

Under the constraint of a limiting velocity imposed by the theory of
relativity, the three distributions share an interesting property in
the temperature dependence. In particular, above a critical
temperature ($T_c =1/\xi_c$), the shape of the function changes from a
nearly Gaussian to bimodal (see Fig.~\ref{distribution1d}).  As a
result, while below the critical temperature the majority of the
particles move at speeds close to the group velocity $U$ (zero in
Fig.~\ref{distribution1d}), above criticality two populations of
particles emerge, with a velocity distribution sharply peaked around
opposite speeds, close to the speed of light.  Here, we show that this
change in the functional dependence of the velocity distribution can
be described in the framework of a transition, which might be either
discontinuous (first order) or continuous (higher order), depending on
the group velocity $U$.  In two and three dimensions, the same
qualitative behavior is observed, where, instead of two peaks, a ring
(in two dimensions) and a hollow sphere (in three dimensions, see
Fig.~\ref{distribution3d}) is obtained at higher temperatures.  In
both cases, at sufficiently high temperatures, the radius in velocity
space reaches the value corresponding to the speed of light (see
Fig.~\ref{distribution3d}).

\begin{figure}
  \includegraphics[width=\columnwidth]{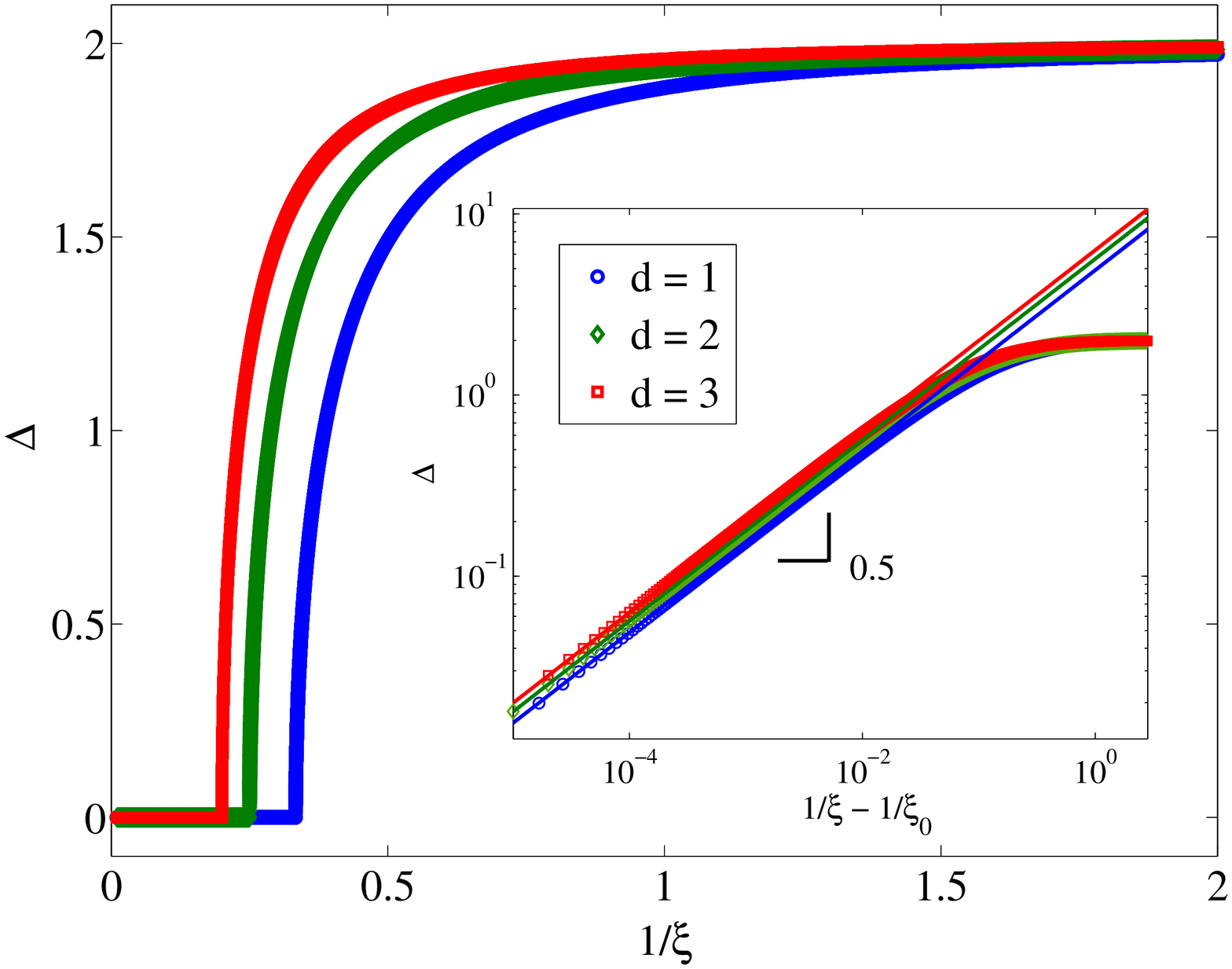}
  \includegraphics[width=\columnwidth]{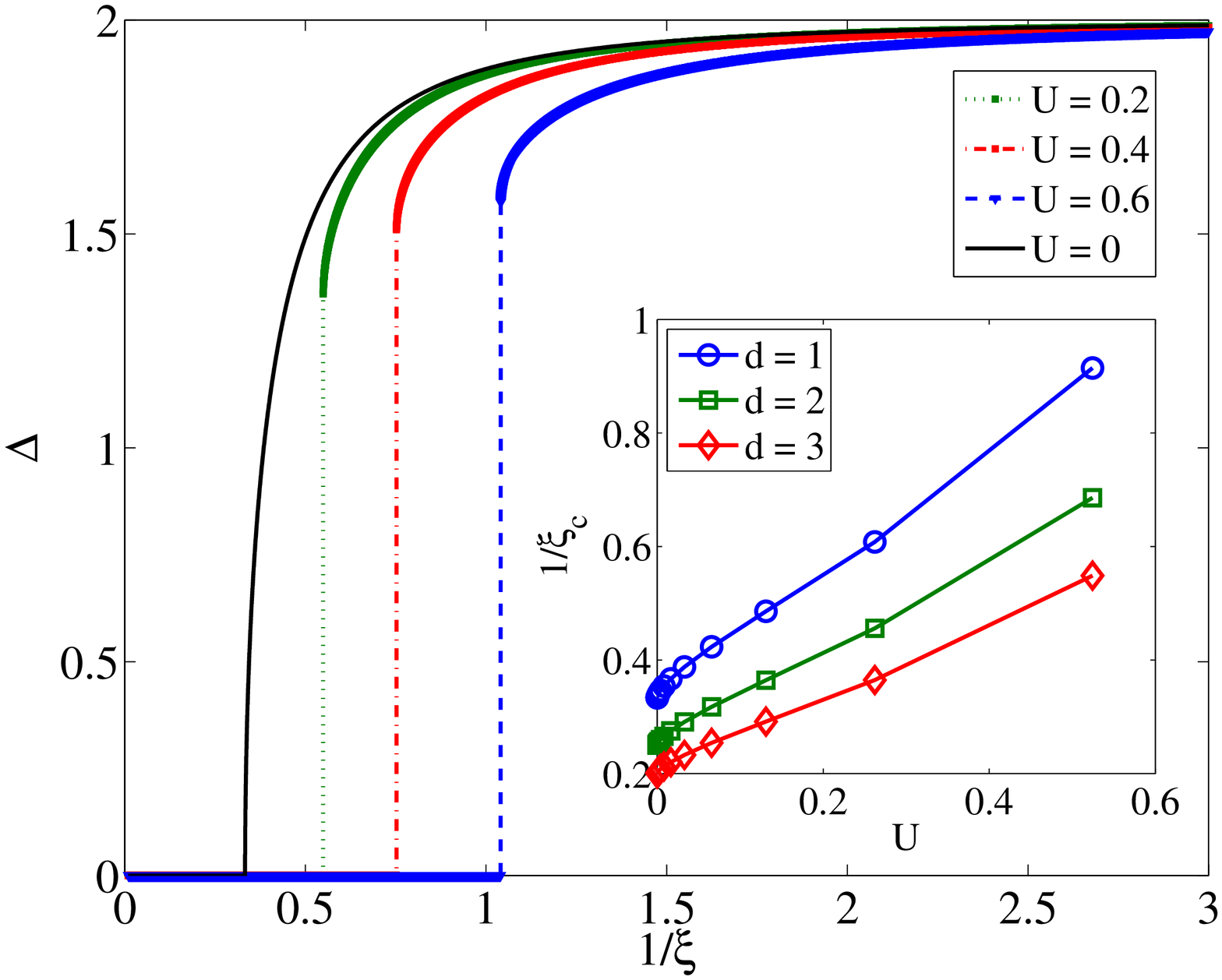}
  \caption{(Top panel) Temperature ($T=1/\xi$) dependence of the order
    parameter $\Delta$, defined as the distance between peaks, for the
    Maxwell-J\"uttner distribution ($\lambda=0$), in different spatial
    dimensions $d$ and group velocity $U=0$. Shown in the inset is a
    double-logarithmic plot of $\Delta$ as a function of the rescaled
    control parameter $1/\xi-1/\xi_c$, where $\xi_0$ stands for the
    transition temperature.  (Bottom panel) Temperature dependence of
    the $\Delta$ for different group velocities. The inset shows the
    transition temperature $T_c=1/\xi_c$ dependence on the group
    velocity $U$.\label{figdelta}}
\end{figure}

To characterize the transition, we introduce an order parameter
defined as the distance $\Delta$ between the peaks, such that
$\Delta=0$ in the single peaked distribution and $\Delta\neq 0$ in the
bimodal one.  Let us begin by considering the case $U= 0$, and measure
the order parameter dependence on the temperature, $T=1/\xi$, as shown
in Fig.~\ref{figdelta}(a).  From this figure, we can appreciate that,
below the critical temperature $T_c=1/\xi_c$, the distribution
function has only one peak, $\Delta=0$, while above $T_c$, a bimodal
profile develops, with the order parameter growing from zero to an
asymptotic value $\Delta = 2$.  In this limit, the width of the
distribution shrinks to zero and the distribution is a superposition
of two Dirac deltas, literally corresponding to a discrete fluid
moving at $\pm c$.

In the inset of Fig.~\ref{figdelta}(a), we analyze the singularity at
$T=T_c$, namely, we plot the order parameter dependence on the
rescaled control parameter $1/\xi-1/\xi_c$.  A continuous transition
is observed, with the order parameter being zero at the critical
temperature and growing according to a power law
$\Delta\sim(1/\xi-1/\xi_c)^{0.5}$ above it.  This exponent corresponds
to the inverse of the exponent $\delta$ in the theory of critical
phenomena \cite{Stanley71}. The same qualitative behavior is observed
in two and three dimensions.  Below, we describe the way the exponent
and critical temperature can be obtained analytically, in the limit $U
\rightarrow 0$.

Due to the fact that the aforementioned transition is driven by the
$\gamma^{d+2}$ pre-factor in the distribution, which is symmetric
around $v=0$ in velocity space, one can, without loss of generality,
calculate $\Delta$ (diameter of the ring, for $d=2$, or of the hollow
sphere, for $d=3$) along the direction $v_x$.  In the limit $U=0$,
from the calculation of the maxima of the distribution,
Eq.\eqref{distribution}, one obtains the trivial solution $v_x = 0$
and two additional ones, corresponding to the solutions of the
algebraic equation,
\begin{equation}
  (2+d) T \lambda + e^{\gamma(v)/T}((2+d)T - \gamma(v)) = 0 \quad .
\end{equation}
Since $\Delta = |v_{x1} - v_{x2}|$, where $v_{x1}$ and $v_{x2}$ are
the two non-zero solutions, we obtain
\begin{equation}
  \Delta = 2\frac{\sqrt{[2+d + W(q)]^2T^2-1}}{T (2 + d + W(q))} \quad ,
\end{equation}
where $q = (2+d)e^{-(2+d)}\lambda$ and $W(x)$ is the Lambert
\mbox{W-function}. From this equation, we can also obtain the critical
temperature as,
\begin{equation}
  T_c = \frac{1}{d+2 + W(q)} \quad ,
\end{equation}
where, in the case of the Maxwell-J\"uttner distribution
($\lambda=0$), i.e., $q = W(q) = 0$, and so $T_c = 1/(d+2)$.  In
general, for this distribution, the temperature dependence of $\Delta$
is given by,
\begin{equation}
  \Delta = \frac{2}{\varepsilon + T_c}\sqrt{\varepsilon(\varepsilon + 2T_c)} \quad ,
\end{equation}
with $\varepsilon=T-T_c$. For $T\gtrsim T_c$, $\Delta \sim \sqrt{8
  \varepsilon/T_c}$, where the exponent is $1/2$, in excellent
agreement with the numerical data in the inset of
Fig.~\ref{figdelta}(a). This exponent depends neither on the spatial
dimension, nor on $\lambda$, i.e., it is the same for the three
distributions. However, the critical temperature $T_c$ depends on
these quantities, and can be obtained analytically for each value of
$d$ and $\lambda$.
\begin{figure}
  \includegraphics[width=\columnwidth]{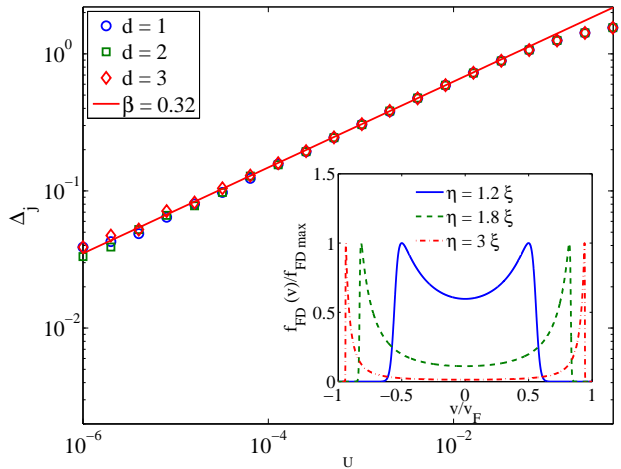}
  \caption{Size of the jump $\Delta_j$ as a function of the group
    velocity $U$ for different dimensions $d$. The quantity $\beta$
    denotes the slope of the best fit. Shown in the inset, is the
    two-dimensional Fermi-Dirac distribution according to
    Eq.~\eqref{fermi} for $\xi = 61.6$ ({\it h}-BN graphene at
    $5$K).\label{figdelta2}}
\end{figure}

Figure~\ref{figdelta}(b) shows the numerical results for the
temperature dependence of $\Delta$ at different group velocities $U$,
for the Maxwell-J\"uttner distribution. For any $U \neq 0$, a jump on
$\Delta$ is observed at the onset of the transition, resembling
thermal first-order transitions. In Fig.~\ref{figdelta2}, we plot the
size of the jump $\Delta_j$ as a function of $U$, where a power law is
obtained $\Delta_j\sim U^\beta$, with $\beta=0.32\pm 0.02$. This
exponent corresponds to the order parameter exponent $\beta$ in the
theory of critical phenomena \cite{Stanley71}. Within the error bars,
the same exponent was obtained for all combinations of $\lambda$ and
$d$, suggesting that $\beta$ is independent of these two parameters.
Nevertheless, as shown in the inset of Fig.~\ref{figdelta}(b), the
transition temperature increases with $U$ and decreases with $d$.

The above transition is not transmitted to the conserved macroscopic
quantities (energy-momentum) and consequently, it is not
straightforwardly observed at the macroscopic level.  Instead, a
microscopic analysis is required.  In order to experimentally detect
the transition, thermal energies of the order of the rest energy of
the gas particles ($\xi \sim 1$) would be required. This could be
achieved, for example, in hot electron plasmas. In particular, it is
of great interest to explore whether this transition can contribute to
a deeper understanding of small condensed matter systems with
potential technological applications. A good candidate in this respect
is graphene. Since its discovery \cite{natletter,Geim1}, graphene has
continued to surprise scientists with an amazing series of spectacular
properties, such as ultra-high electrical conductivity, ultra-low
viscosity to entropy ratio, combination of exceptional structural
strength and mechanical flexibility, and optical transparency. It
consists of literally a single carbon monolayer and represents the
first instance of a truly two-dimensional material (the ``ultimate
flatland''\cite{PhysToday}), where electrons move like massless chiral
particles and their dynamics is governed by the Dirac equation,
following the dispersion relation, $E(\vec{k}) = s \hbar v_F
|\vec{k}|$, where $\vec{k}=(k_x,k_y)$ is the wave vector. The constant
$s = \pm 1$ distinguishes between electrons ($+$) and holes ($-$), and
$v_F$ is the Fermi velocity that plays the same role in graphene as
the speed of light in relativity. This relation implies that carriers
always move at the same Fermi speed, regardless of the Fermi
energy. For simplicity, we will work only with the electronic density
($s=1$), the extension to include holes being straightforward.

In our context, pristine graphene corresponds to the
ultra-relativistic limit, where the velocity distribution function
consists of two Dirac deltas at $\pm v_F$, so that the transition
cannot be observed. However, the electronic spectrum of graphene
changes depending on the substrate. For example, on SiC the energy
spectrum presents a gap of width $2mv_F^2 = 0.26$eV and on {\it h}-BN
(hexagonal boron nitride) a gap of $53$meV
\cite{massive,massive2,massive3}, and can be manipulated by
constructing graphene nanoribbons, where the energy gap depends on the
width of the ribbon \cite{GNR1,GNR2}.

A gap in the spectrum of graphene corresponds to non-zero mass of the
electrons, so that the dispersion relation becomes $E_\alpha(\vec{k})
= \hbar v_F \sqrt{\vec{k}^2 + \alpha^2}$, where $\alpha = mv_F/\hbar$.
With this dispersion relation, the velocity of the electrons can be
calculated as $\vec{v} = (1/\hbar)\partial E(\vec{k})/\partial
\vec{k}$, thus obtaining, $\vec{v} = v_F \vec{k}/\sqrt{\vec{k}^2 +
  \alpha^2}$. Note that the velocity is no longer constant and depends
on the Fermi energy via $\vec{k}$. The electronic density $n$ is
defined by
\begin{equation}
  n = \int f(\vec{k}) \frac{d^2 k}{4\pi^2} \quad ,
\end{equation}
where $f(\vec{k}) = [1+\exp((E(\vec{k})-\mu)/k_B T)]^{-1}$ is the
Fermi-Dirac distribution. To calculate the density of states in
velocity space, we change variables, from the wave to the velocity
vector spaces, obtaining $n = \int D(\vec{v}) f(\vec{v}) d^2 v$, where
$D(\vec{v}) = \alpha^2 \gamma^4/(2\pi)^2 v_F$ is defined as the
density of states (DoS). The Fermi-Dirac distribution function takes
the form,
\begin{equation}\label{fermi}
  f(\vec{v}) = \frac{1}{1 + \exp( \xi \gamma(v) - \eta )} \quad ,
\end{equation}
with $\eta = \mu/k_B T$, and the parameter $\xi$ now defined as $\xi =
mv_F^2/k_B T$. Note that the DoS tends to push all the particles
moving at the Fermi velocity, while the Fermi-Dirac distribution
counters this effect, given the limit imposed by the Fermi speed. At
low temperature, the chemical potential can be approximated by the
Fermi energy \cite{ashcroft}, $E_F$, which can be tuned experimentally
by a gate voltage for low concentration of electrons \cite{Sarma}.
Since low temperatures are required, to avoid electron-phonon
interactions, we characterize the transition of the velocity
distribution functions by changing the Fermi energy, i.e.  $\eta$,
instead of the temperature. With this change in the control parameter
we also have the advantage that the speed at which the Fermi-Dirac
distribution attains its maximum corresponds, to a good approximation,
to the maximum speed at which carriers can move (see inset of
Fig.~\ref{figdelta2}). Therefore, we expect that the transition can be
observed by measuring the thermal or Johnson-Nyquist noise
\cite{johnson, nyquist, noise}. Let us assume that we have a typical
sample of graphene on {\it h}-BN, and the electronic density is
manipulated with an external gate voltage by the relation $n \simeq
\kappa \eta$ \cite{gatedensity}, where $\kappa$ is a proportionality
constant that depends on the electric capacitance of the substrate and
the temperature. Thus, for low carrier concentration, even if there is
no drain voltage, there are current fluctuations $\delta I$ around
zero due to the thermal motion of the electrons in the sample. The
maximum amplitude for these fluctuations can be written as $\delta
I_{\text{max}} = e n v_{\text{max}} l$, where $e$ and $l$ are the
electric charge of the electrons and the cross section of the sample,
respectively. We define the dimensionless maximum amplitude for the
current fluctuations as $\Gamma = 2\delta I_{\text{max}}/e n l v_F$.

\begin{figure}
  \includegraphics[width=\columnwidth]{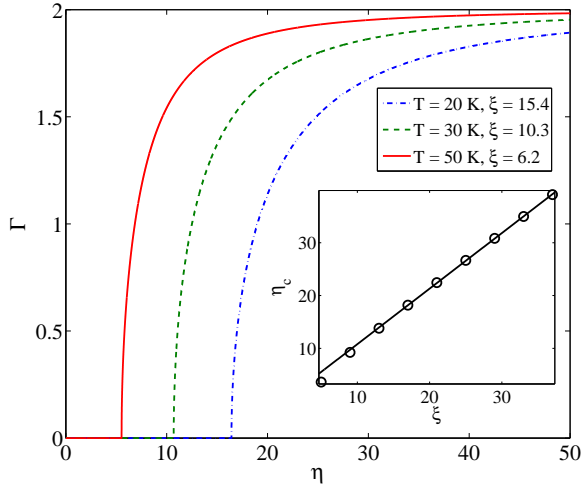}
  \caption{Dimensionless maximum amplitude for the current
    fluctuations $\Gamma$ for gaped graphene on a substrate
    {\it h}-BN as a function of $\eta$. The inset shows the critical
    $\eta_c$, for which the thermal noise emerges, as a
    function of the gap-induced mass, described by $\xi$.  The solid
    line represents the equation $\eta_c = 1.05 \xi + 0.26$.
    \label{peaks_ef}}
\end{figure}

For a fixed low temperature, we can observe from Fig.~\ref{peaks_ef}
that there is a critical $\eta_c$ above which $\Gamma$ emerges. This
critical value increases with decreasing temperature.  We have
considered different temperatures, $T = 20, 30, 50$ K.  Note that the
critical $\eta_c = 1.05 \xi + 0.26$, is almost the same as $\xi$, by a
proportionality constant $1.05$, and the residual is just due to the
non-linear behavior of the curve, close to the critical temperature
where, due to the thermal energy, the thermal noise appears,
regardless of $\eta$. Note that this expression is tantamount to
stating that the Fermi level must be higher than half of the gap
energy in order to have electrons in the conduction band. The exponent
of the continuous transition is also $0.5$, and is independent of the
temperature of the sample in the regime of low temperatures.


Some transport properties in graphene may also depend on the velocity
instead of the momentum, e.g. saturation current in ballistic regime
\cite{drift1, drift2} and mobility of carriers \cite{mobility}.
Therefore, it is plausible to expect that the transition in the
Fermi-Dirac distribution discussed in this Letter, might play an
important role in developing a better understanding of (some of) the
spectacular properties of graphene.  Furthermore, with the aim of
building graphene transistors at nano-size scales, the study of
nanoribbons (GNR) \cite{GNR1,GNR2} has become very popular and,
therefore, the individual carrier dynamics might play a major role in
affecting the transport properties of these devices.

Summarizing, we have shown that the equilibrium distribution for
relativistic particles, no matter whether classical or quantum,
exhibits a transition as the temperature is brought close to the rest
energy (ultra-relativistic limit). This transition is the organized
response of the distribution of particles to the constraint of an
increasing temperature, compatibly with the existence of a limiting
speed for the propagation of physical signals. We have also discussed
conditions under which such transition could potentially be detected
in current and future graphene experiments. In the case of graphene,
we have found that the transition takes place, not only by increasing
the temperature, but also by increasing the Fermi energy at costant
low temperature.

\bibliography{report}

\end{document}